\documentclass[fleqn,usenatbib]{mnras}

\usepackage{newtxtext,newtxmath}

\usepackage[T1]{fontenc}

\DeclareRobustCommand{\VAN}[3]{#2}
\let\VANthebibliography\thebibliography
\def\thebibliography{\DeclareRobustCommand{\VAN}[3]{##3}\VANthebibliography}

\usepackage{graphicx}	
\usepackage{amsmath}	
\usepackage{subcaption}
\usepackage{dsfont}
\usepackage{booktabs}
\usepackage{orcidlink}
\usepackage[colorinlistoftodos, color=lime]{todonotes}

\DeclareMathAlphabet{\mathcal}{OMS}{cmsy}{m}{n} 

\newcommand\bovermat[2]{%
  \makebox[0pt][l]{$\smash{\overbrace{\phantom{%
    \begin{matrix}#2\end{matrix}}}^{#1}}$}#2}

\title[Accounting for Noise and Singularities]{Accounting for Noise and Singularities in Bayesian Calibration Methods for Global 21-cm Cosmology Experiments}

\author[C. J. Kirkham et al.]{Christian J. Kirkham \textsuperscript{\orcidlink{0000-0001-5385-6329}},$^{1,2}$\thanks{E-mail: \href{mailto:cjk55@cam.ac.uk}{cjk55@cam.ac.uk}} 
William J. Handley \textsuperscript{\orcidlink{0000-0002-5866-0445}},$^{1,2}$
Jiacong Zhu \textsuperscript{\orcidlink{0009-0004-8965-0671}},$^{3,4}$
Kaan Artuc \textsuperscript{\orcidlink{0000-0001-8510-5159}},$^{1,2}$
Ian L. V. Roque \textsuperscript{\orcidlink{0000-0003-4874-9371}},$^{5}$\newauthor
Samuel A. K. Leeney \textsuperscript{\orcidlink{0000-0003-4366-1119}},$^{1,2}$
Harry T. J. Bevins \textsuperscript{\orcidlink{0000-0002-4367-3550}},$^{1,2}$
Dominic J. Anstey \textsuperscript{\orcidlink{0000-0003-1742-7417}},$^{1,2}$ and Eloy de Lera Acedo \textsuperscript{\orcidlink{0000-0001-8530-6989}}$^{1,2}$
\\
$^{1}$Astrophysics Group, Cavendish Laboratory, J. J. Thomson Avenue, Cambridge, CB3 0HE, UK\\
$^{2}$Kavli Institute for Cosmology, Madingley Road, Cambridge, CB3 0HA, UK\\
$^{3}$National Astronomical Observatory, Chinese Academy of Sciences\\
$^{4}$School of Astronomy and Space Science, University of Chinese Academy of Sciences\\
$^{5}$SLAC National Accelerator Laboratory, 2575 Sand Hill Road, Menlo Park, California 94025, USA\\\\
}

\date{Accepted XXX. Received YYY; in original form ZZZ}

\pubyear{2025}

\begin{document}
\label{firstpage}
\pagerange{\pageref{firstpage}--\pageref{lastpage}}
\maketitle

\begin{abstract}
Due to the large dynamic ranges involved with separating the cosmological 21-cm signal from the Cosmic Dawn from galactic foregrounds, a well-calibrated instrument is essential to avoid biases from instrumental systematics. In this paper we present three methods for calibrating a global 21-cm cosmology experiment using the noise wave parameter formalisation to characterise a low noise amplifier including a careful consideration of how calibrator temperature noise and singularities will bias the result. The first method presented in this paper builds upon the existing conjugate priors method by weighting the calibrators by a physically motivated factor, thereby avoiding singularities and normalising the noise. The second method fits polynomials to the noise wave parameters by marginalising over the polynomial coefficients and sampling the polynomial orders as parameters. The third method introduces a physically motivated noise model to the marginalised polynomial method. Running these methods on a suite of simulated datasets based on the REACH receiver design and a lab dataset, we found that our methods produced a calibration solution which is equally as or more accurate than the existing conjugate priors method when compared with an estimate of the calibrator's noise. We find in the case of the measured lab dataset the conjugate priors method is biased heavily by the large noise on the shorted load calibrator, resulting in incorrect noise wave parameter fits. This is mitigated by the methods introduced in this paper which calibrate the validation source spectra to within 5\% of the noise floor.
\end{abstract}

\begin{keywords}
instrumentation: interferometers -- methods: data analysis -- cosmology: dark ages, reionization, first stars -- cosmology: early Universe
\end{keywords}



\section{Introduction}

The aim of 21-cm cosmology experiments is to measure the emission from neutral hydrogen (HI) in the periods of the early universe known as the Cosmic Dark Ages, Cosmic Dawn and the Epoch of Reionisation. They detect this emission using the hyperfine transition at a rest frequency of $\nu \approx 1420$ MHz or wavelength $\lambda \approx 21$ cm produced by the HI gas and define a statistical `spin temperature' that is measured relative to the temperature of cosmic microwave background \citep{furlanettoCosmologyLowFrequencies2006}. Using this measurement we can infer the properties of the first galaxies and dark matter in the cosmic dawn \citep{monsalveResultsEDGESHighband2018,monsalveResultsEDGESHighBand2019, bevinsAstrophysicalConstraintsSARAS2022}.

There are several experiments which are designed to measure the global sky-averaged 21-cm signal such as Experiment to Detect the Global EoR Signature (EDGES) \citep{bowmanEmpiricalConstraintsGlobal2008}, Shaped Antenna measurement of the background Radio Spectrum (SARAS) \citep{singhSARASSpectralRadiometer2018,singhDetectionCosmicDawn2022}, Large Aperture Experiment to Detect the Dark Ages (LEDA) \citep{priceDesignCharacterizationLargeaperture2018}, Probing Radio Intensity at high-Z from Marion (PRIZM) \citep{philipProbingRadioIntensity2019}, Mapper of the IGM Spin Temperature (MIST) \citep{monsalveMapperIGMSpin2023} and Radio Experiment for the Analysis of Cosmic Hydrogen (REACH) \citep{deleraacedoREACHRadiometerDetecting2022a}. These experiments use low-frequency radio antennae to detect the 21-cm signal from neutral hydrogen in the Cosmic Dawn and Epoch of Reionisation to place constraints on the physics of the early universe.

The first claimed detection of the global 21-cm signal was made by the EDGES team \citep{bowmanAbsorptionProfileCentred2018}. EDGES found the best fitting profile to be a $0.5^{+0.5}_{-0.2}$ K deep flattened Gaussian centred at $78\pm1$ MHz. This detection raised concerns because of its unusually deep absorption trough and unphysical galactic foreground model \citep{hillsConcernsModellingEDGES2018}, requiring exotic physics to explain the detection \citep{barkanaPossibleInteractionBaryons2018, fengEnhancedGlobalSignal2018}. Another 21-cm global signal experiment, SARAS \citep{nambissanSARASCDEoR2021}, recently placed constraints on the global 21-cm signal, rejecting the EDGES detection with 95.3\% confidence \citep{singhDetectionCosmicDawn2022}.

An alternative explanation for the EDGES signal is uncorrected for systematics in the data \citep{hillsConcernsModellingEDGES2018,singhRedshifted21Cm2019,simsTestingCalibrationSystematics2020}. \cite{hillsConcernsModellingEDGES2018} found evidence of a sinusoidal systematic in the EDGES data, finding the goodness-of-fit improving when a 12.5 MHz sinusoid is removed. This result is consistent with the results of \cite{simsTestingCalibrationSystematics2020}.

In this work we will focus on the REACH experiment \citep{deleraacedoREACHRadiometerDetecting2022a}, a single antenna experiment which is designed to verify the EDGES detection. To do so requires careful characterisation and calibration of systematics to guarantee the sensitivity required to detect the global 21-cm signal, for example \cite{roqueBayesianNoiseWave2021}, \cite{shenBayesianDataAnalysis2022}, \cite{scheutwinkelBayesianEvidencedrivenDiagnosis2022a}, \cite{pattisonModellingHotHorizon2023}, \cite{kirkhamBayesianMethodMitigate2024}, and \cite{cumnerEffectsAntennaPower2024}. In this paper we will present four methods of calibration for 21-cm global experiments using Bayesian methods to characterise the response of the receiver. We will introduce three methods which are similar to the approach used in \cite{roqueBayesianNoiseWave2021}, but with modifications to consider the variation of noise of the calibrated temperatures. In section \ref{s:methods} we will introduce the REACH receiver system and present the calibration methods. In section \ref{s:method_comparison} we will introduce the techniques used to generate a set of simulated data and present a set of lab data, and use both to demonstrate and compare our calibration methods. Finally in section \ref{s:conclusions} we will present our conclusions.

\section{Methods} \label{s:methods}

\subsection{REACH Receiver Calibration}

\begin{figure*}
    \centering
    \includegraphics[width=0.8\linewidth]{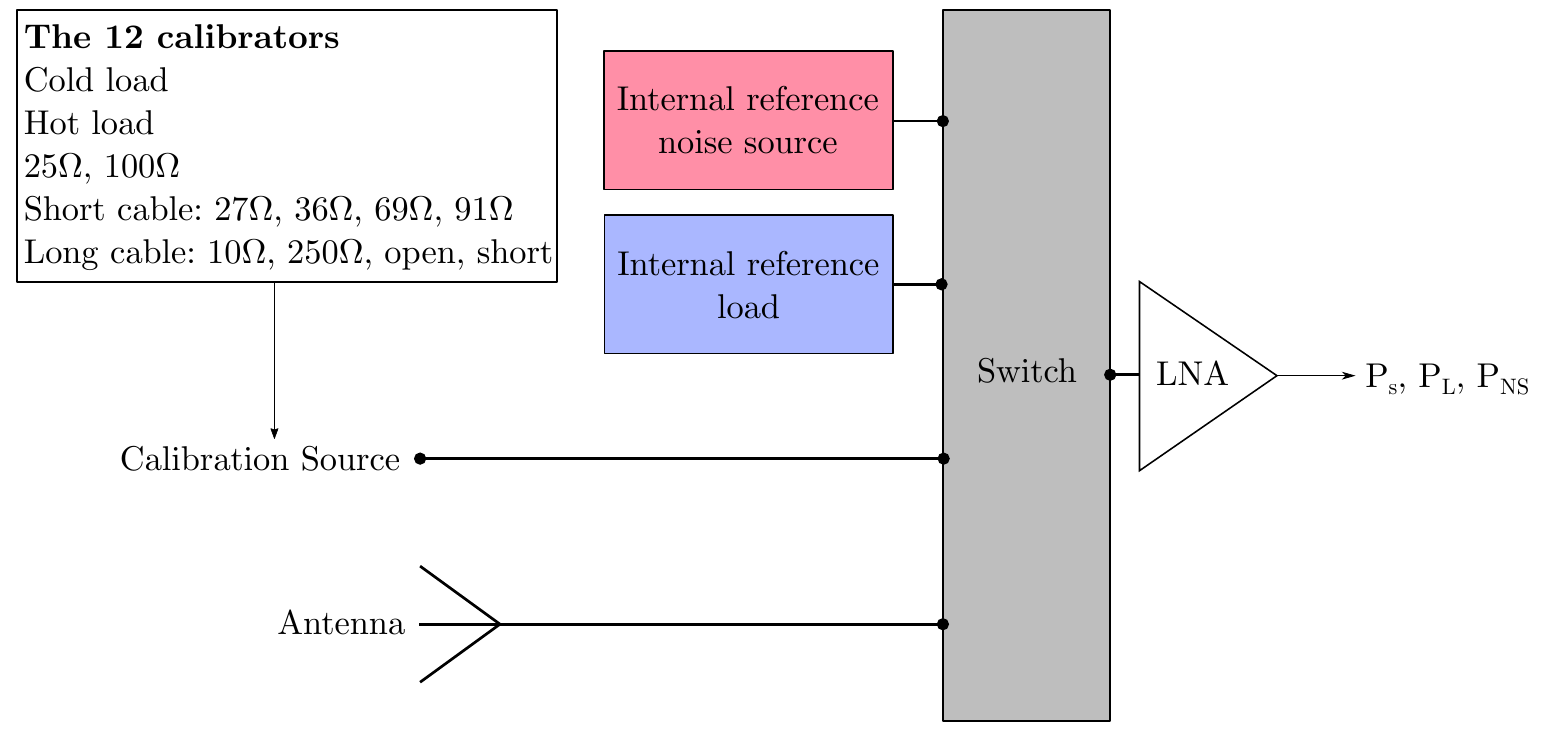}
    \caption{Diagram of the REACH calibration setup using a Dicke switch. Adapted from \protect\cite{roqueBayesianNoiseWave2021}.}
    \label{fig:dicke-switching}
\end{figure*}

To calibrate the instrument the REACH system uses the method of `Dicke switching' which was outlined in \cite{dickeMeasurementThermalRadiation1946} and adapted for global 21-cm experiments in \cite{rogersAbsoluteCalibrationWideband2012}. In order to do this, we measure two reference sources -- a resistive load and a noise source. To account for impedence mismatches in the system we also have twelve calibration sources \citep{meysWaveApproachNoise1978} which were chosen with the aim of maximising the Smith chart coverage, allowing us to measure an unknown source such as the antenna \citep{roqueReceiverDesignREACH2025}. A schematic of this calibration method can be seen in figure \ref{fig:dicke-switching}.

The twelve calibration sources in the REACH receiver are listed below.
\begin{itemize}
    \item An ambient 50 $\Omega$ `cold' load
    \item Ambient 25 $\Omega$ and 100 $\Omega$ loads
    \item A 50 $\Omega$ heated `hot' load at 370K which is connected to a 4 inch cable
    \item 27 $\Omega$, 36 $\Omega$, 69 $\Omega$ and 91 $\Omega$ ambient loads which are connected to a `short' 2 metre cable
    \item 10 $\Omega$, 250 $\Omega$, open and short loads which are connected to a `long' 10 metre cable
\end{itemize}
The antenna is a hexagonal dipole \citep{cumnerRadioAntennaDesign2022} which is connected to a cable approximately one metre in length.

For each calibrator we measure the power spectral density (PSD) of the calibration source, $P_\text{s}$, the PSD of the reference load, $P_\text{L}$, and the PSD of the noise source, $P_\text{NS}$, using a spectrometer after amplification by a low noise amplifier (LNA). We also measure the reflection coefficient of the calibration source connected to the receiver input, $\Gamma_\text{s}$, and the reflection coefficient of the receiver itself, $\Gamma_\text{r}$, by measuring their $S_{11}$s with a vector network analyser (VNA).

Using the \cite{meysWaveApproachNoise1978} noise wave approach of modelling the LNA response we can write the calibrated source temperature, $T_\text{s}$, in terms of the PSDs and reflection coefficients as \citep{monsalveCALIBRATIONEDGESHIGHBAND2017,roqueBayesianNoiseWave2021},
\begin{align} \label{e:full_calibration_eqn}
\begin{split}
    T_\text{NS} \left(\frac{P_\text{s} - P_\text{L}}{P_\text{NS} - P_\text{L}}\right) + T_\text{L} = &T_\text{s} \left[ \frac{1 - |\Gamma_\text{s}|^2}{|1 - \Gamma_\text{s} \Gamma_\text{r}|^2} \right]\\
    &+ T_\text{unc} \left[ \frac{|\Gamma_\text{s}|^2}{|1 - \Gamma_\text{s} \Gamma_\text{r}|^2} \right]\\
    &+ T_\text{cos} \left[ \frac{\mathrm{Re}\left( \frac{\Gamma_\text{s}}{1 - \Gamma_\text{s}\Gamma_\text{r}} \right)}{\sqrt{1 - |\Gamma_\text{r}|^2}} \right]\\
    &+ T_\text{sin} \left[ \frac{\mathrm{Im}\left( \frac{\Gamma_\text{s}}{1 - \Gamma_\text{s}\Gamma_\text{r}} \right)}{\sqrt{1 - |\Gamma_\text{r}|^2}} \right],
\end{split}
\end{align}
where $T_\text{unc}$, $T_\text{cos}$ and $T_\text{sin}$ are functions of frequency known as the noise wave parameters (NWPs) which are fitted for \citep{meysWaveApproachNoise1978}. We also fit for an `effective' load temperature, $T_\text{L}$, and `effective' noise temperature of the internal noise source, $T_\text{NS}$. For brevity we will hereafter refer to all five functions as the `noise wave parameters'. In this work we will be fitting these five functions as polynomials.

Following \cite{roqueBayesianNoiseWave2021}, we can simplify equation \ref{e:full_calibration_eqn} by defining the following terms,
\begin{equation} \label{e:x_matrices}
X_\text{unc} = - \frac{|\Gamma_\text{s}|^2}{1 - |\Gamma_\text{s}|^2},
\end{equation}
\begin{equation} \label{e:X_load}
X_\text{L} = \frac{|1 - \Gamma_\text{s} \Gamma_\text{r}|^2}{1 - |\Gamma_\text{s}|^2},
\end{equation}
\begin{equation}
X_\text{cos} = -\mathrm{Re}\left(\frac{\Gamma_\text{s} }{1 - \Gamma_\text{s}\Gamma_\text{r}} \times \frac{X_\text{L}}{\sqrt{1 - |\Gamma_\text{r}|^2}}\right),
\end{equation}
\begin{equation}
X_\text{sin} = -\mathrm{Im}\left(\frac{\Gamma_\text{s} }{1 - \Gamma_\text{s}\Gamma_\text{r}} \times \frac{X_\text{L}}{\sqrt{1 - |\Gamma_\text{r}|^2}}\right),
\end{equation}
\begin{equation} \label{e:X_noise_source}
    X_\text{NS} = \left(\frac{P_\text{s} - P_\text{L}}{P_\text{NS} - P_\text{L}}\right) X_\text{L},
\end{equation}
which contain all of the measured quantities of our receiver system. We can hence rearrange equation \ref{e:full_calibration_eqn} to get the simple linear calibration equation
\begin{equation} \label{e:simplified_calibration_eqn}
    T_s(\nu) = X_\mathrm{unc} T_\mathrm{unc} + X_\mathrm{cos} T_\mathrm{cos} + X_\mathrm{sin} T_\mathrm{sin} + X_\mathrm{NS} T_\mathrm{NS} + X_\mathrm{L} T_\mathrm{L}.
\end{equation}

Finally we can define the matrices
\begin{equation}
    \bld{X} = \begin{pmatrix}X_\text{unc} & X_\text{cos} & X_\text{sin} & X_\text{NS} & X_\text{L}\end{pmatrix},
\end{equation}
\begin{equation}
    \bld{\Theta} = \begin{pmatrix}T_\text{unc} & T_\text{cos} & T_\text{sin} & T_\text{NS} & T_\text{L}\end{pmatrix}^T,
\end{equation}
which reduces the calibration equation down to the linear equation,
\begin{equation} \label{e:calibration_equation}
    \bld{T}_\text{s} = \bld{X}\bld{\Theta},
\end{equation}
where $\bld{T}_\text{s}$ is a vector of calibrated source temperatures over frequency.

\subsection{Noise Estimation}

In order to have a benchmark by which to test our calibration methods, we estimate the noise on the calibrated temperature by propagating the PSD noise through equation \ref{e:full_calibration_eqn}. Since the S11 noise is orders of magnitude smaller than the PSD noise we make the assumption that the reflection coefficient measurements are noiseless. This means that the only contribution to noise in the final calibrated temperature comes from the measurements of $P_\text{s}$, $P_\text{L}$ and $P_\text{NS}$ in $X_\text{NS}$, equation \ref{e:X_noise_source}.

The noise estimate is as follows,
\begin{align} \label{e:noise_estimate}
\begin{split}
    (\sigma^T_s)^2 = \left(\frac{T^\text{fit}_\text{NS}X_\text{L}}{P_\text{NS} - P_\text{L}}\right)^2& \Bigg(\sigma_{P_\text{s}}^2 + \sigma_{P_\text{L}}^2 - 2\sigma_{P_\text{s}P_\text{L}}\\
    &+ \left(\frac{P_\text{s} - P_\text{L}}{P_\text{NS} - P_\text{L}}\right)^2 (\sigma^2_{P_\text{L}} + \sigma^2_{P_\text{NS}}\\
    &- 2\sigma_{P_\text{L}P_\text{NS}}) - 2\frac{P_\text{s} - P_\text{L}}{P_\text{NS} - P_\text{L}}\sigma_{AB} \Bigg),
\end{split}
\end{align}
where $A = P_\text{s} - P_\text{L}$ and $B = P_\text{NS} - P_\text{L}$. $\sigma_{IJ}$ denotes the covariance of $I$ and $J$. Note that $T^\text{fit}_\text{NS}$ is the fitted valued of $T_\text{NS}$, and that the noise is a frequency dependent quantity. The noise on the PSDs is determined by smoothing the PSD and subtracting the smooth function, leaving the noise as the residual. In order to estimate the noise before fitting, we approximate $T_\text{NS}^\text{fit}$ using the excess noise ratio (ENR) of the noise diode used in the receiver system. In the REACH system we use a noise diode with an ENR of around 6 dB.

\subsection{Bayesian Inference}

To estimate both the noise wave parameters and the parameters describing the noise of each of the calibrated temperatures, we will apply Bayesian inference, a statistical method that reverses conditional probabilities through Bayes' theorem:
\begin{equation}
    P(\bld \theta|\bld D, \mathcal M) = \frac{P(\bld D|\bld \theta, \mathcal M) \cdot P(\bld \theta | \mathcal M)}{P(\bld D | \mathcal M)} = \frac{\mathcal{L}(\theta)\cdot\pi(\theta)}{\mathcal{Z}},
\end{equation}
Here, $\bld \theta$ represents the parameters of the model, $\mathcal M$, that we aim to fit, and $\bld D$ denotes the vector of data points \citep{siviaDataAnalysisBayesian2006}. The term $P(\bld \theta | \mathcal M)$, or $\pi (\bld \theta)$, is the `prior distribution,' which reflects our prior knowledge of the parameter's probability distribution. The `likelihood', $P(\bld D|\bld \theta, \mathcal M)$, or $\mathcal L (\bld \theta)$, is the probability of observing the data given the assumed model and parameters. The `posterior distribution,' $P(\bld \theta|\bld D, \mathcal M)$, or $\mathcal P (\bld \theta)$, represents the probability of the parameters given the data and model, inferred by combining the prior and likelihood distributions. Lastly, $P(\bld D|\mathcal M)$, often denoted as $\mathcal Z$ or the `Bayesian evidence,' can be used to compare different models.

We use a Gaussian likelihood for calibration \citep{roqueBayesianNoiseWave2021} of the form,
\begin{equation} \label{e:gaussian_likelihood}
    \mathcal L = \frac{1}{\sqrt{|2\pi \bld{C}|}}\exp\left\{-\frac{1}{2} (\bld{T}_s - \bld{X\Theta})^T \bld{C}^{-1} (\bld{T}_s - \bld{X\Theta}) \right\},
\end{equation}
where $\bld{C}$ is the covariance matrix which incorporates our calibrator noise parameters.

\subsection{$\Gamma$-Weighted Conjugate Priors}

\cite{roqueBayesianNoiseWave2021} uses a conjugate prior method to quickly evaluate the posterior probability by using a multivariate normal inverse gamma distribution as the prior on the polynomial coefficients. As the likelihood is Gaussian this results in a posterior which is also a normal inverse gamma distribution which can be evaluated analytically from the prior and the likelihood. This is a computationally efficient method for calibration but it makes the assumption that all of the calibration sources have the same calibrator noise parameter, $\sigma$. This assumption is problematic as the noise in the calculated calibrator temperature is proportional to $1 / (1 - |\Gamma_\text{s}|^2)$ (from equations \ref{e:X_load} and \ref{e:noise_estimate}) which can theoretically vary between 1 and $\infty$ for the different calibrators, thereby inflating the radiometric noise in the PSD measurement, $P_\text{s}$.

We also found a serious issue for all inference methods when using an open or short load as in equation \ref{e:x_matrices} all of the $\bld{X}$ matrices are proportional to $1 / (1 - |\Gamma_\text{s}|^2)$, resulting in a singularity in the likelihood for loads with $|\Gamma_\text{s}| \approx 1$. This was highlighted as an issue in \cite{sutinjoMeasureWellSpreadPoints2020} and \cite{priceMeasuringNoiseParameters2023} who show that it is possible to develop methods to mitigate the issues caused by singularities.

In this paper we will use a similar method to \cite{sutinjoMeasureWellSpreadPoints2020} to resolve the issue of singularities due to the $1/ (1 - |\Gamma_\text{s}|^2)$ factor in the $\bld{X}$ matrices by rewriting equation \ref{e:simplified_calibration_eqn} as
\begin{equation} \label{e:gamma_weighted_calibration_eqn}
    \left(1 - |\Gamma_\text{s}|^2\right) \bld{T}_\text{s} = \left(1 - |\Gamma_\text{s}|^2\right) \bld{X\Theta},
\end{equation}
and writing a new calibration equation,
\begin{equation}
    \bld{T}'_\text{s} = \bld{X}'\bld{\Theta},
\end{equation}
where primed quantities represent the new singularity-free quantities, ${\bld{T}'_\text{s} = (1 - |\Gamma_\text{s}|^2) \bld{T}_\text{s}}$, which are finite for all $\Gamma_\text{s}$. The polynomial coefficients, $\bld{\Theta}$, are left unchanged in this new formalisation and we can continue to use the conjugate priors method. Note that in order to find the final calibration solution it is still necessary to calculate the calibrator source temperature,
\begin{equation}
    \bld{T}_\text{s} = \frac{1}{1 - |\Gamma_\text{s}|^2} \bld{T}'_\text{s}.
\end{equation}
As a secondary effect we found that this resulted in all of the calibrated source temperatures exhibiting almost equal RMSE noise levels, allowing the conjugate priors method to fit a single $\sigma$ parameter without biasing the final result. As the noise in the system is radiometric the noise will still vary slightly between calibrators due to temperature variance, but to an extent that it does not significantly bias the final result.

In practice this is equivalent to weighting the calibrators by the factor $(1-|\Gamma_\text{s}|^2)$ in the likelihood. We want to highlight the similarities of this method to the EDGES Bayesian calibration framework \citep{murrayBayesianCalibrationFramework2022} where the authors found it necessary to downweight their open and short cables but noted that their approach is not self-consistent. We believe that this $\Gamma$-weighting method is a physically motivated way of downweighting cables, as open and short cables will have a weight of $(1 - |\Gamma_\text{s}|^2) \sim 0$ while a matched 50 $\Omega$ load will have a weight of $(1 - |\Gamma_\text{s}|^2) \sim 1$. This may suggest that these the open and short cables are not required to calibrate with this method.

\subsection{Marginalised Polynomial Method} \label{s:marg_poly_method}

As an alternative to the conjugate priors method we can fit for each calibrator noise parameter separately using a numerical sampler such as \textsc{PolyChord} \citep{handleyPolychordNestedSampling2015, handleyPolychordNextgenerationNested2015}. In \cite{roqueBayesianNoiseWave2021}, the conjugate priors method used a gradient ascent method to determine the polynomial orders with the highest Bayesian evidence -- a method which is unfeasible with a slow numerical fit and can get stuck in local minima.

In this work we will outline a new method to determine the optimal NWP polynomial order by sampling the polynomial orders as fit parameters. To do so we first introduce two parameter vectors,
\begin{equation}
    \bld{n} = \begin{pmatrix}n_\text{unc} & n_\text{cos} & n_\text{sin} & n_\text{NS} & n_\text{L}\end{pmatrix},
\end{equation}
the vector of polynomial orders for each of the noise wave parameters, and
\begin{equation}
    \bld{\eta} = \begin{pmatrix}\sigma_0& \sigma_1& \dots\end{pmatrix},
\end{equation}
the vector of calibrator noise parameters for each of the twelve calibrators. In this case we are fitting a single monochromatic noise parameter for each calibrator. We can then write the covariance matrix in terms of the calibrator noise parameters as,
\begin{center}
\begin{equation}
    \bld{C} = \mathrm{diag} \begin{pmatrix}\bovermat{N_\nu}{\sigma^2_0& \sigma^2_0& \dots}& \bovermat{N_\nu}{\sigma^2_1& \sigma^2_1& \dots}\end{pmatrix},
\end{equation}
\end{center}
where $N_\nu$ is the number of frequency channels in the data. We can now say that the likelihood is a function of the three parameter vectors $\mathcal L (\eta, \bld n, \bld \Theta)$.

Similar to approaches taken by EDGES and HERA \citep{simsBayesianPowerSpectrum2017, monsalveResultsEDGESHighband2018, tauscherGlobal21Cm2021, murrayBayesianCalibrationFramework2022}, we can exploit the linearity of the likelihood in equation \ref{e:gaussian_likelihood} and analytically marginalise over the polynomial parameter vector, $\bld \Theta$, in order to speed up the sampling. This also simplifies the likelihood by removing the need for transdimensional sampling methods \citep{heeBayesianModelSelection2016, kroupaKernelMeanNoisemarginalised2023} when sampling the NWP polynomial orders.

We do this by setting a Gaussian prior on the polynomial coefficients,
\begin{equation} \label{e:gaussian_prior}
    \pi(\bld{\Theta}) = \frac{1}{\sqrt{|2\pi \bld{\Sigma}_\pi|}}\exp\left\{-\frac{1}{2} (\bld{\Theta} - \bld{\mu}_\pi)^T \bld{\Sigma}_\pi^{-1} (\bld{\Theta} - \bld{\mu}_\pi) \right\},
\end{equation}
where we have the mean, $\bld{\mu}_\pi$, and the prior covariance matrix, ${\bld{\Sigma}_\pi = \sigma_\pi^2 \bld{I}}$. Using this prior to marginalise over the polynomial parameters we hence get the marginal log likelihood,
%
\begin{align}
\begin{split}\label{e:marginal_likelihood}
    \log \mathcal{L}(\bld{\eta},\bld{n}) = &\frac{1}{2} \log\left|\frac{\bld{\Sigma}_P}{\bld{\Sigma}_\pi}\right| - \frac{1}{2} (\bld{\mu}_P - \bld{\mu}_\pi)^T \bld{\Sigma}_\pi^{-1} (\bld{\mu}_P - \bld{\mu}_\pi) \\ &- \frac{1}{2}\log |2\pi\bld{C}| - \frac{1}{2}(\bld{T} - \bld{X}\bld{\mu}_P)^T \bld{C}^{-1}(\bld{T} - \bld{X}\bld{\mu}_P),
\end{split}
\end{align}
with
\begin{align}
\begin{split}
    \bld{\Sigma}_P^{-1} &= \frac{1}{\sigma_\pi^2} \bld{I} + \sum_i \frac{1}{\sigma_i^2} \bld{X}_\mathrm{i}^T \bld{X}_\mathrm{i}, \\
    \bld{\mu}_P &=  \frac{1}{\sigma_\pi^2}\bld{\Sigma}_P\bld{\mu}_\pi + \sum_i \frac{1}{\sigma_i^2} \bld{\Sigma}_P \bld{X}_\mathrm{i}^T \bld{T}_\mathrm{s,i},   
\end{split}
\end{align}
where $i$ denotes the $i$th calibration source. Here, $\bld{C}$ is a function of $\bld{\eta}$ and the sizes of $\bld{\Sigma}_P$ and $\bld{\mu}_P$ depend on $\bld{n}$. Once the posteriors of $\bld \eta$ and $\bld n$ have been sampled with \textsc{PolyChord} we can then find the posterior on the polynomial coefficients as draws from a multivariate normal distribution,
\begin{equation}
    \bld \Theta \sim \mathcal{N} (\bld \mu_P, \bld \Sigma_P),
\end{equation}
as per equation \ref{e:gaussian_prior}.

We set an exponential prior, $n \sim \mathrm{Exp}(5)$, on the polynomial orders and a log uniform prior, $\sigma_i \sim \mathrm{LogUniform}(10^{-4},0.1)$ K on the calibrator noise parameters. The width of the Gaussian prior on the polynomial coefficients is set to $\sigma_\pi = 10$ K, with the prior mean for the zeroth polynomial coefficients informed by the least squares solution \citep{roqueReceiverDesignREACH2025}. For this work we set the zeroth order coefficient means to 270 K, 175 K, 40 K, 1120 K and 300 K for $T_\text{unc}$, $T_\text{cos}$, $T_\text{sin}$, $T_\text{NS}$ and $T_\text{L}$ respectively. These priors are motivated based on a visual inspection of a preliminary least squares fit to the noise wave parameters \citep{roqueReceiverDesignREACH2025} and provide a good estimate of the true values. Higher order polynomial coefficients have a prior mean of 0 K.

We can also use this method with the $\Gamma$-weighted calibration equation from equation \ref{e:gamma_weighted_calibration_eqn} with no changes to the method. This can be useful if singularities are causing numerical instabilities when fitting.

\subsection{Marginalised Polynomial Method with Noise Model}

One major benefit of the marginalised polynomial method is that it allows us to use arbitrary models for the noise in our likelihood by modifying the covariance matrix, $\bld C$. In this work we will construct the covariance matrix using the frequency-dependent estimated noise from equation \ref{e:noise_estimate}, defining this new matrix as $\tilde{\bld C}$.

Since the noise estimate has a $T_\text{NS}$ term, care must be taken with the likelihood to preserve the linearity of the likelihood necessary to marginalise over the polynomial coefficients. In this work we will assume $T_\text{NS}$ to be a zeroth order polynomial, motivated by inspection of the fits from other polynomial methods. In order to marginalise over the other noise wave parameters, we rearrange equation \ref{e:full_calibration_eqn}, and define the following terms,
\begin{equation}
    \tilde{T}_\text{s} = T_\text{s} - X_\text{NS}T_\text{NS},
\end{equation}
\begin{equation}
    \tilde{\bld{X}} = \begin{pmatrix}X_\text{unc} & X_\text{cos} & X_\text{sin} & X_\text{L}\end{pmatrix},
\end{equation}
\begin{equation}
    \tilde{\bld{\Theta}} = \begin{pmatrix}T_\text{unc} & T_\text{cos} & T_\text{sin} & T_\text{L}\end{pmatrix}^T,
\end{equation}
\begin{equation}
    \tilde{\bld{n}} = \begin{pmatrix}n_\text{unc} & n_\text{cos} & n_\text{sin} & n_\text{L}\end{pmatrix}.
\end{equation}
The likelihood is hence
\begin{equation} \label{e:gaussian_likelihood_noise_model}
    \mathcal L = \frac{1}{\sqrt{|2\pi \tilde{\bld{C}}|}}\exp\left\{-\frac{1}{2} (\tilde{\bld{T}}_s - \bld{\tilde{X}\tilde{\Theta}})^T \tilde{\bld{C}}^{-1} (\tilde{\bld{T}}_s - \bld{\tilde{X}\tilde{\Theta}}) \right\},
\end{equation}
which can be marginalised as before, resulting in the marginal likelihood $\mathcal{L}(\tilde{\bld{n}}, T_\text{NS})$, as in equation \ref{e:marginal_likelihood}. Note that we no longer sample over the calibrator noise parameters as this has been calculated using our fit value of $T_\text{NS}$ and equation \ref{e:noise_estimate}, thereby reducing the total number of parameters that we must numerically sample over to just five.

\subsection{Cable Corrections}

The set of REACH calibrators include nine sources which are connected to the LNA with cables. In a real system we would expect the cables to have a different temperature, $T_\text{c}$, to the calibrator source temperature which will introduce losses and reflected power from the cable. To correct for this, we calculate the realised gain of the cable,
\begin{equation}
    G_\text{c} = \frac{|S_{21}|^2 (1 - |\Gamma_\text{r}|^2)}{|1 - S_{11} \Gamma_\text{r}|^2(1 - |\Gamma_\text{s}|^2)},
\end{equation}
where $S_{11}$ and $S_{21}$ are the forward S-parameters of the cable \citep{monsalveCALIBRATIONEDGESHIGHBAND2017,roqueReceiverDesignREACH2025}. This can then be used to calculate an effective source temperature,
\begin{equation}
    T^\text{eff}_\text{s} = G_\text{c} T_\text{s} + (1-G_\text{c})T_\text{c},
\end{equation}
which is used to calibrate the source and fit the noise wave parameters. Once the source has been calibrated we can rearrange this expression to get the final source temperature,
\begin{equation}
    T_\text{s}^\text{final} = \frac{1}{G_\text{c}} (T^\text{fit}_\text{s} + (G_\text{c}-1)T_\text{c}).
\end{equation}
For $\Gamma$-weighted calibration methods we must also weight the cable temperature as $T_\text{c}' = (1 - |\Gamma_\text{s}|^2) T_\text{c}$ when correcting $T_\text{s}'$.

\section{Method Comparison} \label{s:method_comparison}

\subsection{Testing with Simulated Mock Data} \label{s:mock_data}

In order to test this method we will generate five sets of mock data intended to simulate the REACH receiver system to varying levels of complexity following the methods used in \cite{sunCalibrationError21centimeter2024}. These mock datasets are produced using lab $S_{11}$ and $S_{21}$ measurements of the LNA and simulated noise parameters \citep{priceNewTechniqueMeasure2023}. The five datasets are as follows:

\begin{itemize}
    \item \textbf{Dataset 1:} The antenna is simulated as a flat 5000 K spectrum with the system gain, $g=1$. The temperatures of the cables are set to the temperature of the sources.

    \item \textbf{Dataset 2:} The antenna is simulated as a flat 5000 K spectrum with the system gain, $g=1$. The temperatures of the cables are set to 317 K.

    \item \textbf{Dataset 3:} The antenna is simulated as a flat 5000 K spectrum with a realistic value of the system gain, $g=|S_{21}|^2$. The temperatures of the cables are set to 317 K.

    \item \textbf{Dataset 4:} The antenna is simulated as a power law spectrum with a realistic value of the system gain, $g=|S_{21}|^2$. The temperatures of the cables are set to the temperature of the sources.

    \item \textbf{Dataset 5:} The antenna is simulated as a power law spectrum with a realistic value of the system gain, $g=|S_{21}|^2$. The temperatures of the cables are set to 317 K.
\end{itemize}

In all datasets we do not model the antenna as being connected to a cable. We set the temperature of the hot load to 370 K, the cold load to 320 K and all other sources are randomly selected from a Gaussian distribution centred around 320 K with a standard deviation of 20 K.

The power law spectrum in datasets 4 and 5 is generated using the EDGES log-polynomial fit to the galactic foreground power, \citep{bowmanAbsorptionProfileCentred2018}
\begin{align}
\begin{split}
    T_\text{ant} = &1801.4\left(\frac{\nu}{\nu_0}\right)^{-2.5} + 116.7\left(\frac{\nu}{\nu_0}\right)^{-1.5} -289.8\left(\frac{\nu}{\nu_0}\right)^{-0.5} \\&+ 144.8 \left(\frac{\nu}{\nu_0}\right)^{0.5} - 22.9\left(\frac{\nu}{\nu_0}\right)^{1.5},
\end{split}
\end{align}
where $\nu_0 = 75$ MHz, generated over the range $110 \leq \nu \leq 140$ MHz to match the band of the lab dataset in section \ref{s:real_data}. We do not consider antenna chromaticity effects, assuming an achromatic antenna gain.

For each mock dataset we will then compare the marginalised polynomial method introduced in this paper with the original conjugate priors method \citep{roqueBayesianNoiseWave2021} and the $\Gamma$-weighted conjugate priors method.

\subsection{Testing with Lab Data} \label{s:real_data}

Finally we will test the method on lab measurements of a real system which is designed to be a simplified version of REACH system. The data was measured with an integration time of 5 minutes and we used the data in the range $110 \leq \nu \leq 140$ MHz. This system is made up of seven sources, listed below.

\begin{itemize}
    \item An ambient 50 $\Omega$ `cold' load
    \item An ambient 90 $\Omega$ load at the end of a 0.37 m cable
    \item An ambient 100 $\Omega$ load
    \item A 50 $\Omega$ heated `hot' load heated to approximately 360 K
    \item An ambient short load
    \item 0.3 m and 2 m open-terminated cables
\end{itemize}

In order to validate the calibration and simulate an antenna we leave the 90 $\Omega$ ambient load out of the calibration and compare the calibrated temperature with the measured temperature of the load. This will be referred to as the `validation source'. In this case all sources are plugged directly into the VNA and LNA, avoiding the need to use switches which will cause signal loss. We do not perform any cable temperature gradient corrections and treat the loads and cables as sources.

\subsection{Results and Discussion}

\begin{figure*}
    \centering
    \includegraphics[width=0.8\linewidth]{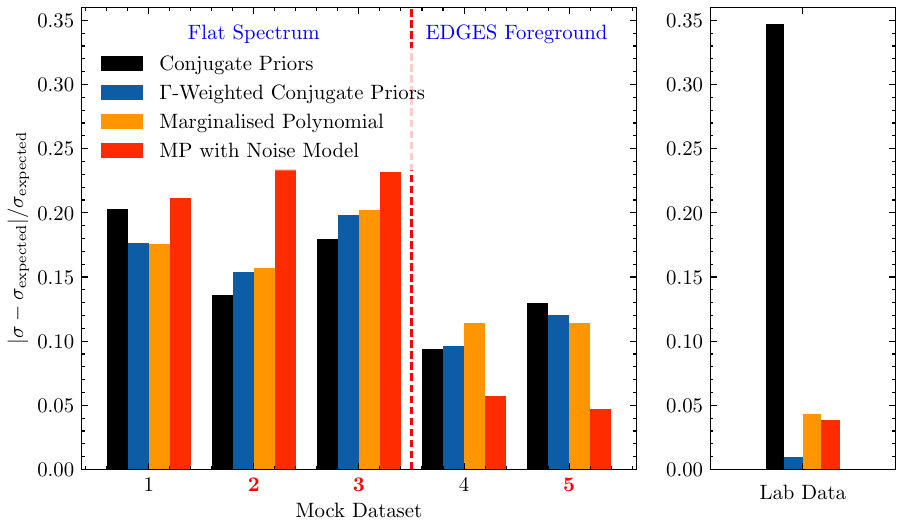}
    \caption{Summary of the three calibration methods tested on the five mock datasets and the lab dataset. Here we plot the fractional difference between the calculated noise of the calibrated validation source solution and the expected noise of the validation source. Mock datasets highlighted in red indicate datasets which include cable temperature gradients. It can be seen that while the methods perform equally for the mock datasets, we see the methods presented in this paper outperform the existing conjugate priors method on a lab dataset.}
    \label{fig:rmse_mock_data}
\end{figure*}

The results of running the four calibration methods on the five mock datasets and the lab dataset can be seen in figure \ref{fig:rmse_mock_data}. We use the fractional error between the RMSE of the residuals of the calibrated validation source temperature and the estimated RMSE value derived from equation \ref{e:noise_estimate} using the true value of $T_\text{NS}$ as a figure of merit. The plot is separated into two, with the datasets which have a flat 5000 K antenna on the left and datasets with a EDGES foreground spectrum on the right.

For all mock datasets we see that all four methods perform comparably, giving antenna solutions whose RMSEs are within 25\% of the noise level. In general we see that the flat spectrum datasets have a higher error, since the absolute calibration error scales with the antenna temperature. Since the flat antenna is 5000 K across the frequency range we see this larger error at all frequencies. In comparison, the EDGES foreground datasets which ranges from 650 K to 350 K across the band have a much lower absolute error overall.

Since the largest reflection coefficient in the mock datasets is of order $|\Gamma_\text{s}| \sim 0.8$, this is far enough from $1$ that the conjugate priors method does not get significantly affected by the noise biasing issues discussed. This produces calibrated temperatures which are very similar to those of the other methods. However, we find negative values of $T_\text{unc}$ which are unphysical as this is the magnitude of the reflected receiver noise. As this results in a good solution for the antenna, we believe this is a result of degeneracies in the noise wave parameter solutions that the three other methods are able to break, but this requires further investigation.


\begin{figure*}
    \centering
    \includegraphics[width=0.8\linewidth]{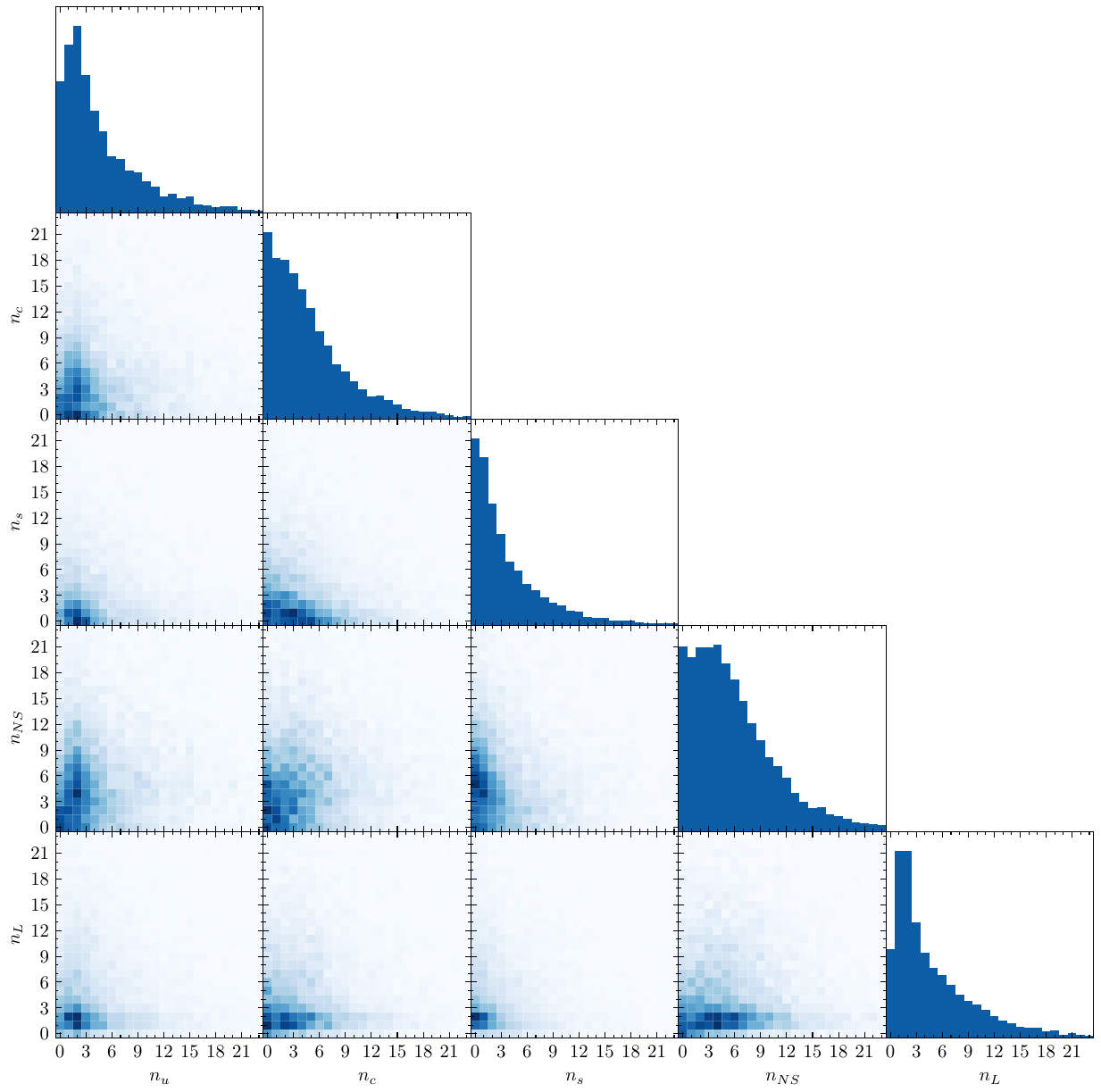}
    \caption{Corner plot of the marginalised polynomial order posterior probabilities for the lab dataset. As the higher orders are disfavoured, we can conclude that the noise wave parameters are not being overfitted by the marginalised polynomial method.}
    \label{fig:poly_order_triangle_plot}
\end{figure*}

In contrast to the mock datasets we see that the three new methods presented in this paper greatly outperform the conjugate priors method for the lab dataset. We note that these methods perform better on the lab data than the mock data because shorter integration times result in higher noise on the lab data PSDs, hiding some residual structure. A corner plot showing the posterior distribution of the polynomial orders can be seen in figure \ref{fig:poly_order_triangle_plot}. All of the posteriors peak at low order polynomials, demonstrating that the methods are avoiding overfitting over the noise wave polynomials. The degree to which high order polynomials are disfavoured can be tuned by adjusting the parameter of the exponential prior.

\begin{figure*}
    \centering
    \includegraphics[width=0.9\linewidth]{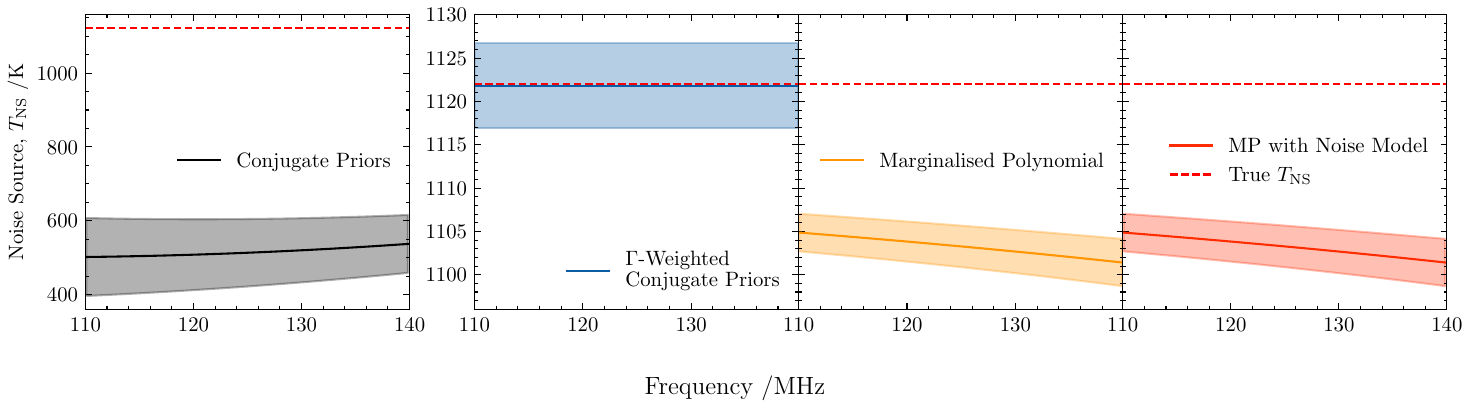}
    \caption{Comparison of the fitted polynomial to $T_\text{NS}$ for each calibration method for the lab dataset. The existing conjugate priors method greatly underestimates the noise source temperature while the methods presented in this paper are closer to the true value. Despite the improvement, the marginalised polynomial methods still slightly underestimate the value of $T_\text{NS}$.}
    \label{fig:TNS_method_comparison}
\end{figure*}

We also find that the $\Gamma$-weighted conjugate priors method performs comparably well to the marginalised polynomial method. This is due to the weighting factor normalising the noise in the calibrated temperature reducing the bias resulting from the same calibrator noise level assumption. In addition we find that the weighting factor increases the numerical stability of the method when fitting the lab data as the short load has a reflection coefficient close to unity.

The reason the conjugate priors method struggles to calibrate the lab data is that the singularities and noise differences bias the fit to the noise source temperature, $T_\text{NS}$. This can be seen in the functional posteriors for $T_\text{NS}$ in figure \ref{fig:TNS_method_comparison}, where the conjugate priors method has a much lower $T_\text{NS}$ value than the other three methods. As a result, as shown in equation \ref{e:noise_estimate}, the noise on the calibrated temperature is much lower than the expected value. This is a key point we wish to highlight --  the smallest RMSE noise value possible is not the goal of calibration, as if it is smaller than the estimated value then it is likely that you have fitted your noise wave parameters incorrectly. 

Since in equation \ref{e:full_calibration_eqn} $P_\text{s}$ is multiplied by $T_\text{NS}$, incorrectly fitting the noise source temperature will result in the 21-cm signal amplitude being scaled incorrectly, just as the noise is scaled. For example, an overestimate of $T_\text{NS}$ will result in a global 21-cm signal depth which is larger than reality. In addition, incorrectly fitting the other noise wave parameters will introduce residual structure in the final calibration solution.

The three new methods presented in this work recover a more realistic $T_\text{NS}$ value of approximately 1100 K for the lab dataset, however we note that the marginalised polynomial methods still slightly underestimate its value. The noise diode used for the lab measurements has an ENR of 6 dB or 800 K, which when added to the ambient temperature, $T_\text{L}$, of approximately 300 K results in the expected $T_\text{NS}$ of around 1100 K. We cannot verify the other noise wave parameters, $T_\text{unc}$, $T_\text{sin}$, or $T_\text{cos}$ independently but, as discussed, we require $T_\text{unc} > 0$.

\begin{figure*}
    \centering
    \includegraphics[width=0.9\linewidth]{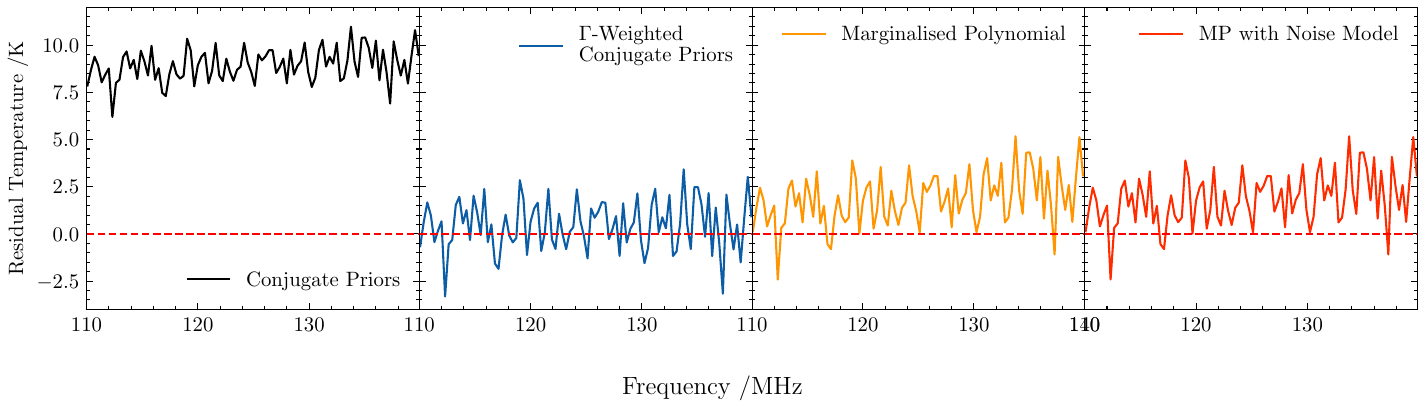}
    \caption{Comparison of the difference between the calibrated validation source spectrum and the measured validation source temperature for each calibration method for the lab dataset. While there is little residual structure in all of the spectra, the absolute calibration is best for the $\Gamma$-weighted conjugate priors method, indicative of the method's ability to constrain $T_\text{NS}$ accurately. The noise level of the residuals here are close to what we expect for the short integration time of the lab dataset.}
    \label{fig:ant_res_method_comparison}
\end{figure*}

We can see the residuals on the validation source in figure \ref{fig:ant_res_method_comparison}. While the methods do not show significant structure we can see there is a visible difference in the absolute offset which can be attributed to the inaccurate fits to $T_\text{NS}$ as demonstrated in figure \ref{fig:TNS_method_comparison}. We also note that the noise level of the residuals are close to what we expect, taking into account the short five minute integration. In order to get the noise down to science-ready noise we would require further integration.

Finally, we wish to highlight that the marginalised polynomial methods and $\Gamma$-weighted conjugate priors method successfully calibrate lab data to within 5\% of the theoretical noise limit, demonstrating the power of these Bayesian methods. In particular we have shown that these methods are more robust than the previous conjugate priors method, even when faced with loads which introduce singularities into the system of equations. Further improvements could be made by improving the fitting of the noise model used in the marginalised polynomial which, for example, assumes that the S11 measurements are noiseless.

\section{Conclusions} \label{s:conclusions}

In this paper we introduced three new Bayesian methods for receiver calibration of global 21-cm experiments and compare them with the conjugate priors method introduced in \cite{roqueBayesianNoiseWave2021} using a suite of simulated REACH datasets and a lab measured dataset. We benchmarked all three methods by finding the difference between the measured RMSE noise of the validation source residuals and the estimated value of the noise.

The lab dataset tests showed the $\Gamma$-weighted conjugate priors method outperformed the original conjugate priors method owing to the noise normalisation and increased numerical stability due to the mitigation of the effects of singularities. We also find that the marginalised polynomial method introduced in this paper also outperforms the conjugate priors method since the ability to fit separate noise parameters for the calibrators adds extra degrees of freedom and prevents the calibration result from being biased by high-noise sources such as a shorted cable. Furthermore, the marginalised polynomial method introduces the freedom to choose an arbitrary noise model which further improves the fit to the lab dataset when a physically motivated covariance matrix is used.

We also highlight that the goal of calibration is not just to have the smallest RMSE -- although lower RMSE does imply less residual structure -- but rather that the RMSE should be as close to the noise estimate as possible. A measured RMSE which is lower than the estimate likely implies that you have incorrectly fitted your noise wave parameters.

\section*{Acknowledgements}

CJK was supported by Science and Technology Facilities Council grant number ST/V506606/1. WJH was supported by a Royal Society University Research Fellowship. SAKL was supported by the European Research Council and the UKRI. HTJB acknowledges support from the Kavli Institute for Cosmology Cambridge and Kavli Foundation. DJA was supported by Science and Technology Facilities Council grant number ST/X00239X/1. EdLA was supported by Science and Technology Facilities Council grant number ST/V004425/1. We would also like to thank the Kavli Foundation for their support of REACH. The authors thank the Science and Technology Facilities Council grant number EP/Y02916X/1 for supporting the REACH project.

\section*{Data Availability}
 
The data that supported the findings of this article will be shared on reasonable request to the corresponding author.



\bibliographystyle{mnras}
\bibliography{21cm_Cosmology}




\bsp	
\label{lastpage}
\end{document}